\UseRawInputEncoding
\documentclass[aps,twocolumn,prm,superscriptaddress,longbibliography]{revtex4-1}
\usepackage{times,amsmath,amssymb,graphicx,multirow}
\usepackage[colorlinks=true,citecolor=blue,urlcolor=blue,anchorcolor=blue,linkcolor=blue,pdfauthor=author]{hyperref}

\begin{document}
\title{Strain-driven valley states and phase transitions in Janus VSiGeN$_4$ monolayer}

\author{Pengyu Liu}
\affiliation{Institute for Computational Materials Science, Joint Center for Theoretical Physics (JCTP), School of Physics and Electronics, Henan University, Kaifeng, 475004, China}

\author{Siyuan Liu}
\affiliation{Institute for Computational Materials Science, Joint Center for Theoretical Physics (JCTP), School of Physics and Electronics, Henan University, Kaifeng, 475004, China}

\author{Minglei Jia}
\affiliation{Institute for Computational Materials Science, Joint Center for Theoretical Physics (JCTP), School of Physics and Electronics, Henan University, Kaifeng, 475004, China}

\author{Huabing Yin}
\affiliation{Institute for Computational Materials Science, Joint Center for Theoretical Physics (JCTP), School of Physics and Electronics, Henan University, Kaifeng, 475004, China}

\author{Guangbiao Zhang}
\affiliation{Institute for Computational Materials Science, Joint Center for Theoretical Physics (JCTP), School of Physics and Electronics, Henan University, Kaifeng, 475004, China}

\author{Fengzhu Ren}
\email{f.z.ren@henu.edu.cn}
\affiliation{Institute for Computational Materials Science, Joint Center for Theoretical Physics (JCTP), School of Physics and Electronics, Henan University, Kaifeng, 475004, China}

\author{Bing Wang}
\email{wb@henu.edu.cn}
\affiliation{Institute for Computational Materials Science, Joint Center for Theoretical Physics (JCTP), School of Physics and Electronics, Henan University, Kaifeng, 475004, China}

\author{Chang Liu}
\email{cliu@vip.henu.edu.cn}
\affiliation{Institute for Computational Materials Science, Joint Center for Theoretical Physics (JCTP), School of Physics and Electronics, Henan University, Kaifeng, 475004, China}
\date{\today}

\begin{abstract}
The interplay between topology and valley degree of freedom has attracted much interest  because it can realize new phenomena and applications. Here, based on first-principles calculations, we demonstrate intrinsically valley-polarized quantum anomalous Hall effect in monolayer ferrovalley material: Janus VSiGeN$_4$, of which the edge states are chiral-spin-valley locking. Besides, a small tensile or compressive strain can drive phase transition in the material from valley-polarized quantum anomalous Hall state to half-valley-metal state. With the increase of the strain, the material turns into ferrovalley semiconductor with valley anomalous Hall effect. The origin of phase transition is sequent band inversion of V d orbital at K valley. Moreover, we find that phase transition causes the sign reversal of Berry curvature and induces different polarized light absorption in different valley states. Our work provides an ideal material platform for practical applications and experimental exploration of the interplay between topology, spintronics, and valleytronics.
\end{abstract}

\maketitle

\section{Introduction}

The valley, local maximum or local minimum of electronic band structure in momentum space, provides valley degree of freedom to encode and manipulate information analogous to charge and spin \cite{1,2,3},
the related field is called valleytronics. For utilizing the valley index as an information carrier, the key is how to break the energy degeneracy of the valley in the valleytronics material, which can create carries imbalance between different valleys to be detected \cite{4}.
The experimental discovery of two-dimensional (2D) transition metal dichalcogenides (TMDs) has greatly promoted the development of valleytronics \cite{5,6,7,8,9,10,11,12,13,14,15,Guansan1,Guansan2}.
It is found that optical pumping can produce valley polarization in MoS$_2$ based on the valley-contrasting optical selection rules \cite{6,7}, a finding that can be applied to valley optoelectronic devices \cite{16,17}.
Unfortunately, due to the short lifetime of the carries, the valley will depolarize quickly and cause information loss. On the other hand, some experiments reported that Zeeman effect of external magnetic field can break the energy degeneracy of valleys in TMD \cite{11,12}.
However, the efficiency of magnetic field is low ($0.1\sim0.2$ meV/T), thereby it is difficult to preserve a suitable valley splitting value by magnetic field in electronic equipment. Besides, it was found that magnetic doping and magnetic proximity effect can induce valley polarization \cite{18,19,20}.
Especially, a larger than 300 meV valley splitting in MoTe$_2$ can be induced by magnetic substrate of EuO \cite{21}.
Yet, the magnetic substrate submerges the valley physics of host materials, and magnetic doping usually increases impurity scatterings between different valleys. Valley polarization induced by above ways is volatile. Thus, it is important to find intrinsic valley polarization.

Recently, valleytronic material have been extended to 2D ferromagnetic system named ferrovalley \cite{22,23,24,25,26,27,28,29,30,31,32,33,34,35}.
When both time-reversal and inversion symmetry are broken, ferrovalley materials can produce spontaneous valley polarization and realize anomalous valley Hall effect (AVHE) \cite{22,25}, and these properties can be well applied to data process or data storage.
Importantly, out-of-plane magnetization is a necessary condition for achieving spontaneous valley polarization \cite{27,31,37}. However, most of the discovered ferrovalley materials have in-plane magnetization \cite{26,27,29,MLVS},
and thus require difficult modulation of the magnetic easy axis. On the other hand, some novel quantum states were discovered in ferrovalley materials, such as topological state and half-valley-metal (HVM) state \cite{30,33}.
The multiple valley states not only enrich valley degree of freedom and quantum states but also bring intriguing physical properties, such as valley-polarized quantum anomalous effect (VQAHE) and selective absorption of polarized light \cite{23,32,33}.
Though highly valuable, ferrovalley material with both modulate multiple valley states and out-of-plane magnetization is quite rare.

In this work, we predict a ferrovalley material of Janus VSiGeN$_4$ monolayer based on density-functional theory (DFT). We find that this monolayer can not only maintain valley characteristics, but also has large valley polarization and strain-induced phase transition. Our results show that monolayer VSiGeN$_4$ is a ferromagnetic semiconductor, and it has out-of-plane magnetization with a Curie temperature of 113K. Interestingly, the material possesses a VQAHE with chiral-spin-valley locking edge states. Besides, it becomes HVM with 100$\%$ valley polarization under tiny tensile strain of 0.1$\%$ or compressive strain of 0.32$\%$. By continuing to increase the strain, the material will transform into ferrovalley semiconductor (FVS), which shows an AVHE. Moreover, the polarized light absorption and the sign of berry curvature located in the K$+$ and K$-$ valley can also be modulated by applying strain. Our work thus enriches materials with valley-related multiple Hall effect and provides a platform for research valleytronics, spintronics, and topology.

\begin{figure}[htbp]
	\centering
	\includegraphics[width=1.0\linewidth]{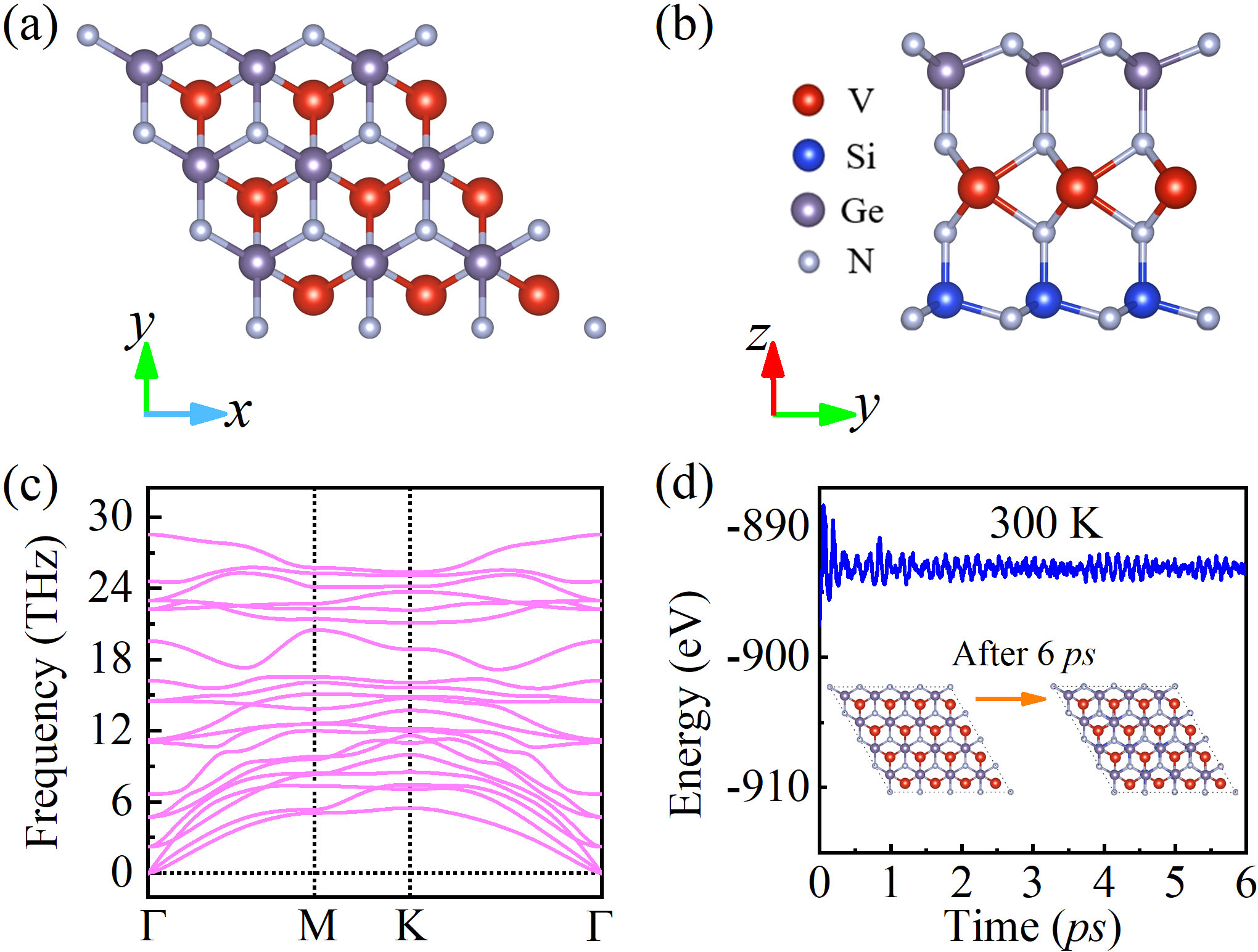}
	\caption{(a) and (b) Top and side views of ML VSiGeN$_4$, respectively. (c) Phonon dispersion spectrum and (d) energy variation during the 6 $ \emph{ps}$ AIMD simulation at 300K. Insets in (d) display the initial and final structures of ML VSiGeN$_4$ in the MD simulation.}
	\label{fgr:1.jpg}
\end{figure}

\section{Computational Methods}
All the first-principles calculations were performed using the projector-augmented wave (PAW) method \cite{paw}, which is embedded in the Vienna $\emph{ab initio}$ Simulation Package (VASP) \cite{vasp}. The exchange and correlation interaction is described by the Predew-Burke-Ernzerhof (PBE) functional with the general gradient approximation (GGA) \cite{GGA}. The spin-orbit coupling (SOC) effect is included in the calculations. We use $ \Gamma$-centered k-meshes of 15$\times$15$\times$1 for the Brillouin-zone (BZ) sampling and 500 eV cutoff energy for the plane wave basis. The convergence criterion for energy and force is chosen as 10$^{-6}$ eV and 0.01 eV/\AA, respectively. A vacuum region of 20~\AA~is taken to eliminate the interactions between periodic layers. The GGA+U method is adopted to describe the strong correlated correction of V-3d orbitals and the U parameter is chosen as 3 eV \cite{23,26,35}. The $\emph{ab initio}$ molecular dynamic (AIMD) simulation is performed by using a 4$\times$4$\times$1 supercell to determin thermal stability. Phonon spectrum is calculated by using the density functional perturbation theory (DFPT) with a 4$\times$4$\times$1 supercell. The edge states and Chern number were calculated by using the WANNIER90 package \cite{wannier} and WANNIERTOOLS package \cite{wanniertools}. The Monte Carlo (MC) simulations are performed on a 2D 16$\times$16 supercell based on the standard Metropolis-Hasting algorithm, and we use 640000 sweeps to get reasonable result.

\section{STRUCTURE STABILITIES AND MAGNETISM}

The Janus monolayer (ML) VSiGeN$_4$ has a triangular lattice structure with space group P3M1, and the optimized lattice constant is a $=$ b $=$ 2.96 \AA. As shown in  Fig.~\ref{fgr:1.jpg}(a) and Fig.~\ref{fgr:1.jpg}(b), the lattice has no space inversion symmetry, and the structure is built up by septuple atomic layers in the sequence of N-Ge-N-V-N-Si-N, which can be viewed as VN$_2$ layer sandwiched by a Ge-N and Si-N layer. To confirm the stability of structure, the phonon spectra and molecular dynamic (MD) simulation are shown in Fig.~\ref{fgr:1.jpg}(c) and Fig.~\ref{fgr:1.jpg}(d), respectively. The absence of imaginary frequencies in phonon spectra confirms the dynamics stability. After 6 $ \emph{ps}$ MD simulation at 300K, no destruction is observed in ML VSiGeN4, which indicate the thermal stability under 300K. Besides, we calculated the formation energy of ML VSiGeN$_4$: $E_f=(E_{total}-E_{Si}-E_V-E_{Ge}-4E_N)/n$.
Where E$_{total}$ represent the total energy of ML VSiGeN$_4$ in the primitive cell, E$_{Si}$, E$_V$, E$_{Ge}$, and E$_N$ is the energy per atom of Si, V, Ge and N in their most stable phases, respectively, and n is the number of atoms in the primitive cell. The calculated value of E$_f$ is about $-$0.602 eV/atom, implying that ML VSiGeN$_4$ has a good thermodynamical stability against the elemental phases. Therefore, we expect that ML VSiGeN$_4$ can be synthesized experimentally, for example by substituting atoms in ML MoSi$_2$N$_4$ \cite{Hong}.

Next, the magnetic properties of ML VSiGeN$_4$ is investigated. We calculate the total energies of different magnetic states: ferromagnetic (FM), nonmagnetic (NM), antiferromagnetic (AFM1 and 120$^{\circ}$ AFM) structures, which are shown in Fig. S1. The results show that the total energy of FM structure is the lowest, 76 meV per unit cell lower than the AFM1 state, 351 meV per unit cell lower than NM state. Besides, the spin configuration of 120$^{\circ}$ AFM is unstable and transforms into FM state, which is similar to 2H-VSSe and LaBrI \cite{28,36}. Thus, FM state is the ground state, which can be interpreted by the Goodenough-Kanamori-Anderson (GKA) rules \cite{wuyaxuan,GKA1,GKA2,GKA3}. In ML VSiGeN$_4$, the V-N-V bonding angle is 91.8$^{\circ}$, close to 90$^{\circ}$. According to the GKA rules, such a structure configuration favors FM coupling. To determine the easy magnetization axis, we calculated magnetic anisotropy energy (MAE), which is defined as E$_M=$ E$_{[100]}-E_{[001]}$, where E$_{[100]}$ and E$_{[001]}$ represent the total energy of system with the spin orientation of V atoms along out-of-plane and in-plane, respectivley. The E$_M$ is 72 $ \mu$eV per unit cell, which means that ML VSiGeN$_4$ is out-of-plane magnetization. Such magnetization direction is necessary to realize the spontaneous valley polarization.

To study the magnetic stability of ML VSiGeN$_4$, we estimated the Curie temperature (T$_C$) by using the classical Metropolis MC simulations based on the Heisenberg model:
$H=-\sum_{<ij>}J_{ij}S_{i}\cdot S_{j}-D{(S_{iz})^2}$.
Where $S_i$ is the normalized spin vector on site $i$, $J_{ij}$ is the exchange interaction strength between sites $i$ and $j$, $D$ is the single-site MAE, $S_{iz}$ represents component of $S$ along $z$ orientation, and $ \lvert{S}\rvert = \frac{1}{2}$ for V in 2D VSiGeN$_4$. Because the moment of the V atom is around 1 $\mu_B$. With $J=$ 51.49 meV and $D=$ 72 $\mu$eV, the result of MC simulation shows that the value of T$_C$ is about 113 K (Fig. S1(d)), which is much higher than CrI$_3$ (45 K) and Cr$_2$Ge$_2$Te$_6$ bilayer (30 K) \cite{CrI$_3$,CrGeTe}.

\begin{figure}[htbp]
	\centering
	\includegraphics[width=1.0\linewidth]{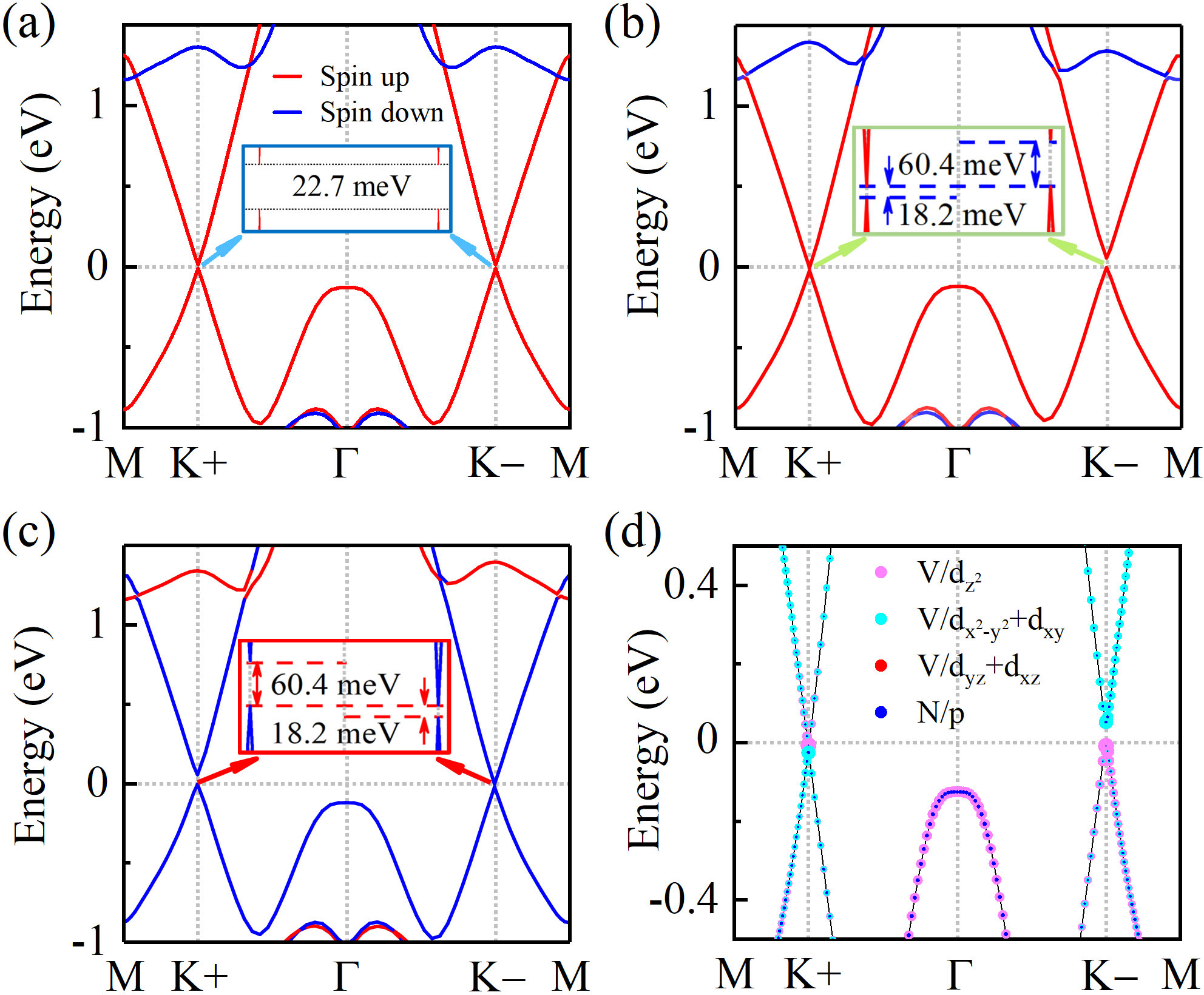}
	\caption{The spin polarized band structures of ML VSiGeN$_4$ (a) without SOC; (b) and (c) with SOC for magnetic moment of V along the positive and negative z direction, respectively. (d) Orbital-resolved spin polarized band structure with SOC in ML material. Spin projections along the out-of-plane direction are indicated by red (positive direction, spin-up) and blue (negative direction, spin-down) line for VSiGeN$_4$ in (b) and (c).}
	\label{fgr:2.jpg}
\end{figure}
\section{ELECTRONIC STRUCTURES AND STRAIN-DRIVEN VALLEY STATES}

The spin polarized band structure of the ML VSiGeN$_4$ without considering SOC is shown in Fig.~\ref{fgr:2.jpg}(a). It can be seen that spin-down and spin-up channels split significantly, and there is only spin-up states around the Fermi level (E$_F$). The material is a semiconductor with a small direct band gap of 22.7 meV. Both the valence band maximum (VBM) and the conduction band minimum (CBM) locate at the high symmetry point K. The degenerate valleys at the K$+$ and K$-$ point are nonequivalent due to the break of the inversion symmetry. Therefore, ML VSiGeN$_4$ is a valleytronic material. Similar to the previously reported ferrovalley materials, when considering both SOC and magnetic exchange interaction, ML VSiGeN$_4$ spontaneously  enables a valley polarization without any additional tuning. As shown in Fig.~\ref{fgr:2.jpg}(b), the SOC lifts the energy degeneracy of K$+$ and K$-$ valley (the energy of K$-$ valley is higher than K$+$ valley for both valence band and conduction band) with valley polarization of 60.4 meV (18.2 meV) at conduction (valence) band. This polarization value is larger than previously those of reported materials, such as VSi$_2$P$_4$ (49.4 meV) \cite{23}, LaBr$_2$ (33meV) \cite{27}, Cr$_2$Se$_3$ (18.7meV) \cite{24} and TiVI$_6$ (22meV) \cite{37}. More importantly, the influence of irrelevant bands for valley properties is minimum in the energy range of ($-0.2\sim1$ eV). The linear band dispersion around K$-$ and K$+$ valley show a high Fermi velocity of about $0.31\times{10}^6~{ms}^{-1}$, which has the same order of magnitude as graphene. The valley polarization can be tuned by an external magnetic field. As shown in Fig.~\ref{fgr:2.jpg}(c), the valley polarization and spin state are flipped by reversing the magnetization of 
V atoms, so that all K valley states now occupy the spin-down channel and K$-$ valley state has a lower energy than K$+$ valley. The tunable large valley polarization, large Fermi velocity, and the clean single spin state around E$_F$, endow VSiGeN$_4$ with more advantages than most valleytronics materials in practical applications. 

The different values of valley polarization between K$+$ and K$-$ valley can be attributed to the different orbital contributions. As shown in Fig.~\ref{fgr:2.jpg}(d), at K$+$ valley, the CBM is contributed by $d_{z^2}$ orbital of V and the VBM is contributed by $d_{x^2-y^2}+d_{xy}$ orbital of V, while the situation at K$-$ valley is reversed. Since the magnetic quantum number ($m_z$) of $d_{z^2}$ orbital is 0, the basis functions composed of this orbital is the same at K$+$ and the K$-$ point. While basis functions composed of $d_{x^2-y^2}$ and $d_{xy}$ orbital is different at K$+$ and the K$-$ point, which can be chosen as: $\left|\psi\right\rangle$=$ \frac{1}{\sqrt{2}}(\left|d_{x^2+y^2}\right\rangle+i\tau\left|d_{xy}\right\rangle )$. Here $\tau=\pm1$ indicates the valley index at the $K\pm$ point \cite{23}. Therefore, introducing SOC into a system with neither time-reversal symmetry nor space-reversal symmetry will result in different energy eigenvalues at K$+$ point and K$-$ point with different basis functions, leading to the occurrence of valley polarization.
\begin{figure}[htbp]
	\centering
	\includegraphics[width=1\linewidth]{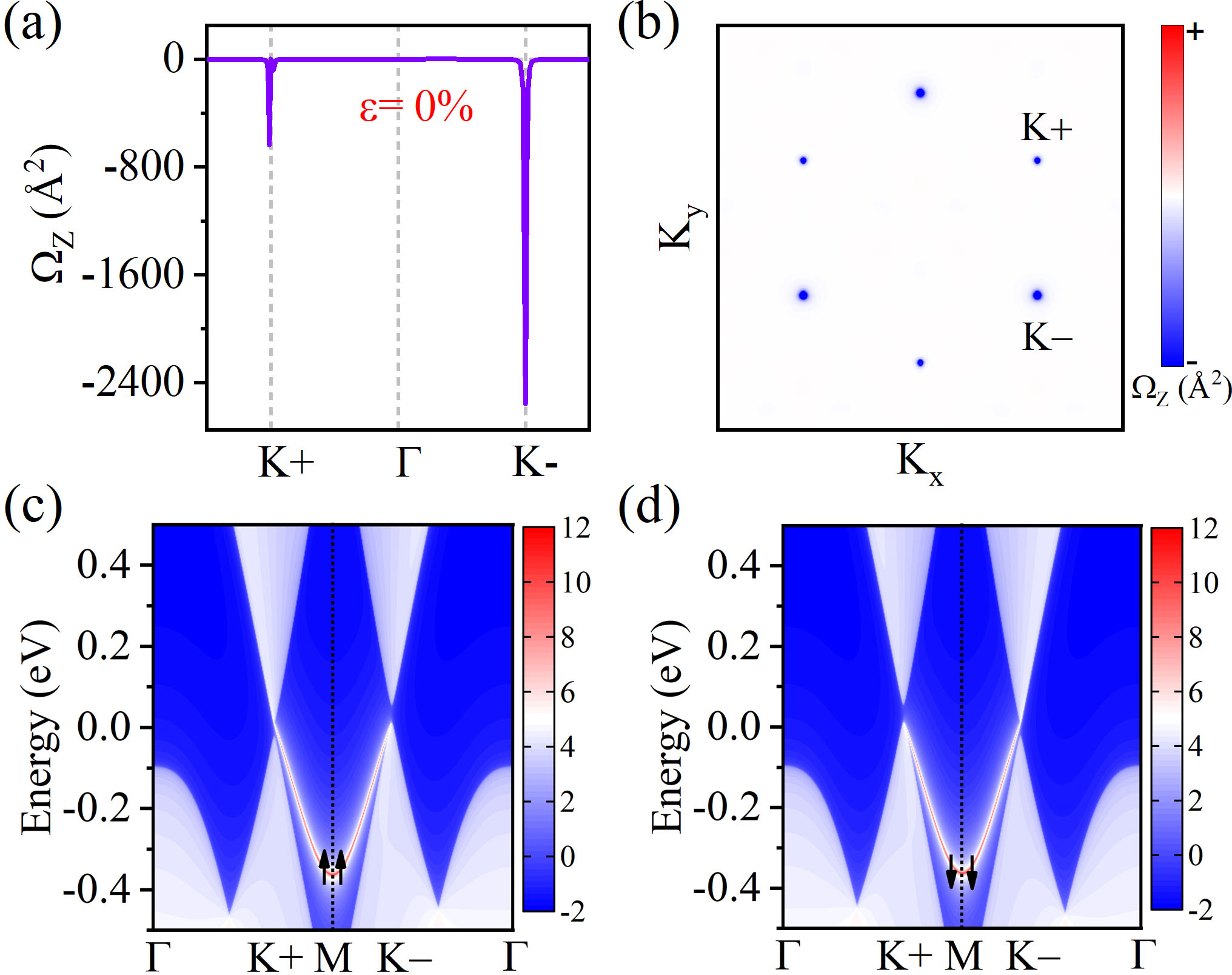}
	\caption{(a) and (b) The distribution of Berry curvature along the high symmetry path and along the whole BZ in strain-free ML VSiGeN$_4$, respectively. (c) and (d) The edge state for ML VSiGeN$_4$ with magnetic moment along positive and negative $z$ direction, respectively.}
	\label{fgr:3.jpg}
\end{figure}

The opposite orbital order for the K$-$ and K$+$ valleys imply that ML VSiGeN4 has topological properties, such as VQAH effect. This anomalous transport phenomenon is related to the Berry curvature $\Omega(k)$. For a 2D systems, $\Omega(k)$ only has $z$ component $\Omega_z(k)$. It can be calculated by \cite{38}:
\begin{align*}
\Omega_z(k)=-\sum_n\sum_{n\neq n^{'}}f_n\frac{2Im\langle\psi_{nk}|\upsilon_x|\psi_{n^{'}k}\rangle\langle\psi_{n^{'}k}|\upsilon_y|\psi_{nk}\rangle}{(E_n-E_n^{'})^2},
\end{align*}
where $n$ and $n^{'}$ are band indexes; $f_n$ is the equilibrium Fermi-Dirac distribution function for the $n$-th band at a k point; $\upsilon_{x(y)}$ is the velocity operator along the $x(y)$ direction; 
\begin{figure*}[htbp]
	\centering
	\includegraphics[width=0.8\linewidth]{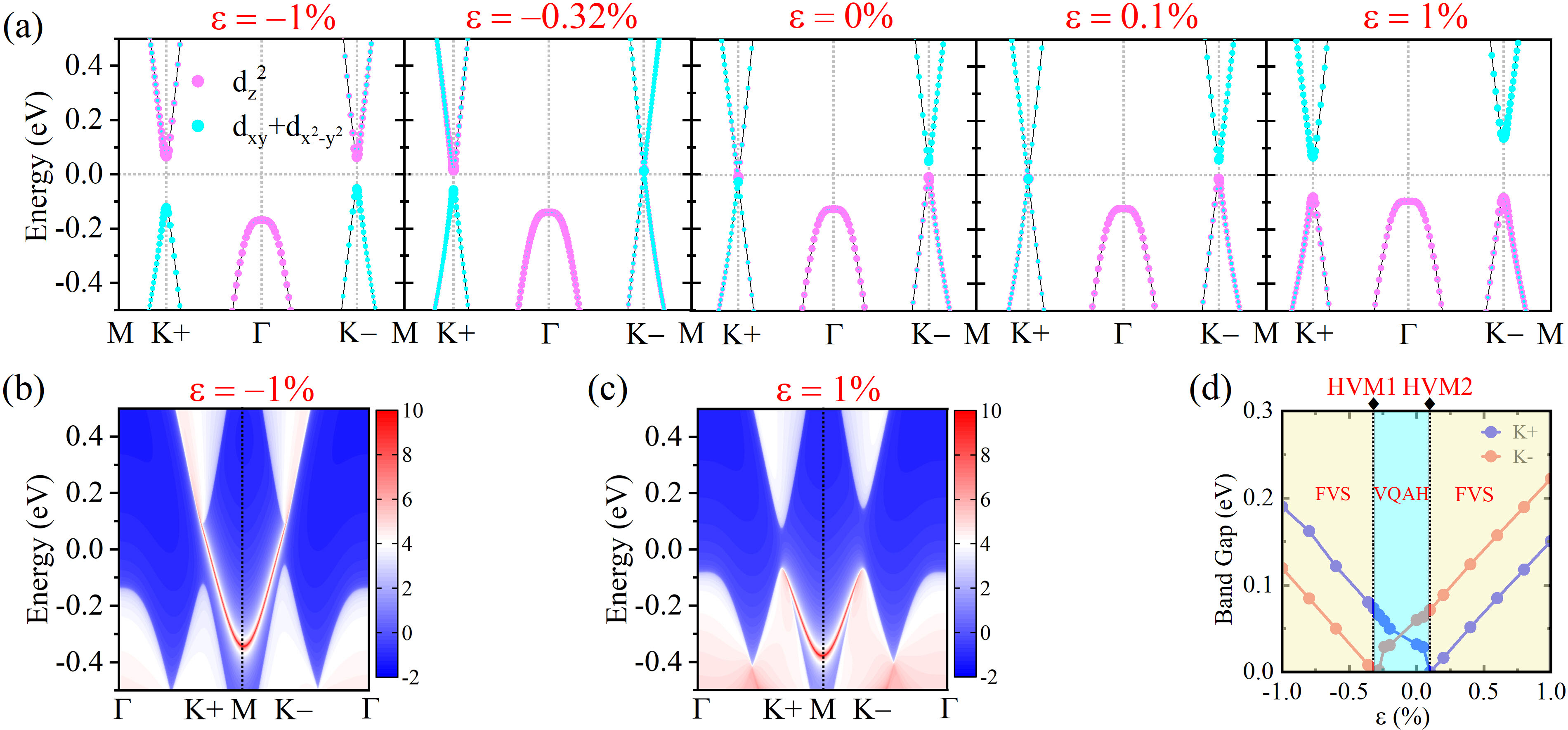}
	\caption{(a) The evolution of orbital-projected band with SOC under different biaxial strain, respectively. (b) and (c) The edge state for ML VSiGeN$_4$ under $\varepsilon=-1\%$ and $\varepsilon=1\%$, respectively. (d) The diagram of topological phase transitions and the variation of the band gaps at K valley with different strain.}
	\label{fgr:4.jpg}
\end{figure*}
$k$ is the wave vector; $\psi_{nk}$ is the Bloch wave function with eigenvalue $E_n$ and $E_{n^{'}}$.

In Fig.~\ref{fgr:3.jpg}(a) and Fig.~\ref{fgr:3.jpg}(b), we show the $\Omega_z(k)$ distribution of ML VSiGeN$_4$ along the high symmetry path and contour map in the 2D Brillouin zone (BZ), respectively. One clearly observes that the values of $\Omega_z(k)$ are all zero except the region around K$+$ and K$-$ valleys. More importantly, the sign is same and the magnitudes are different from two valleys. With such distribution of $\Omega_z(k)$, the material may have a nonzero Chern number and edge states protected by the nontrivial topological property, which result in a VQAH effect. We have evaluated the Chern number (C) of ML VSiGeN$_4$ via formula: $C=\frac{1}{2\pi}\int_{BZ}^{\ }{d^2k{\Omega}_z(k)}$. The result shows that the Chern number is 1. Then, we calculated the edge states as shown in Fig.~\ref{fgr:3.jpg}(c), it can be seen that a topological nontrivial chiral edge state connecting conduction bands (K$+$ valley) and valence bands (K$-$ valley) is clearly visible. Due to CBM and VBM all dominated by spin-up states, the edge state is also spin-up with 100$\%$ spin polarization. Similar to the reversal of valley polarization under external magnetic field, when the magnetization of V is reversed, the edge state will change accordingly. As shown in Fig.~\ref{fgr:3.jpg}(d), it becomes spin-down state and connect conduction bands (K$+$ valley) and valence bands (K$-$ valley) with opposite chiral. Besides, the  conductive edge state of spin-up in the global band gap mainly locate at K$+$ valley, the location of edge state change from K$+$ valley to K$-$ valley when spin flipping, 
which revealed a behavior of the chiral-spin-valley locking for the edge state \cite{30}. Thus, above properties indicate that ML VSiGeN$_4$ features an intrinsically VQAH effect.

The electronic properties of low-dimensional materials can be generally tuned by strain \cite{Guansan3,Liu1,Liu2,Liu3,Liu4,Liu5}. Here, we introduce biaxial strain in ML VSiGeN$_4$ and study the effect of strain on the band structure. The biaxial strain strength is defined as $\varepsilon=\frac{b-b_0}{b_0}$, where $b$ and $b_0$ are the strained and unstrained lattice constant, respectively. The result is shown in Fig.~\ref{fgr:4.jpg}(a) and Fig. S2. It shows that the band gap is obviously changed under different strain. Specially, when $\varepsilon=-$0.32$\%$ or 0.1$\%$, the material transforms into HVM state with 100$\%$ valley polarization and 100$\%$ spin polarization. In such a state, one valley is metallic and the other one is semiconductor with same spin-up state. The gapless valley shows a Dirac cone-like linear dispersion with a Fermi velocity of $3.58\times10^5~{ms}^{-1}$. Changing the strain can open a gap, and the HVM state is thus disappeared. As the strain increases, the band gap of the two valleys also increases gradually, whereas the large valley polarization always remains (see Fig.~\ref{fgr:4.jpg}(d)). Specially, the valley polarization only appears at valence band for $-$1$\%$ strained VSiGeN$_4$, while in the case of 1$\%$ strain, only the conduction band has different energy for K$+$ and K$-$ valley. This phenomenon also originates from the change of orbital composition, as discussed earlier. Only $d_{x^2-y^2}+d_{xy}$ orbital has different energy eigenvalues at different valley, and this orbital constitute the valence band for $\varepsilon=-1\%$ and the conduction band for $\varepsilon=1\%$, respectively. Thus, leading to different valley polarization behaviors.
\begin{figure*}[htbp]
	\centering
	\includegraphics[width=0.8\linewidth]{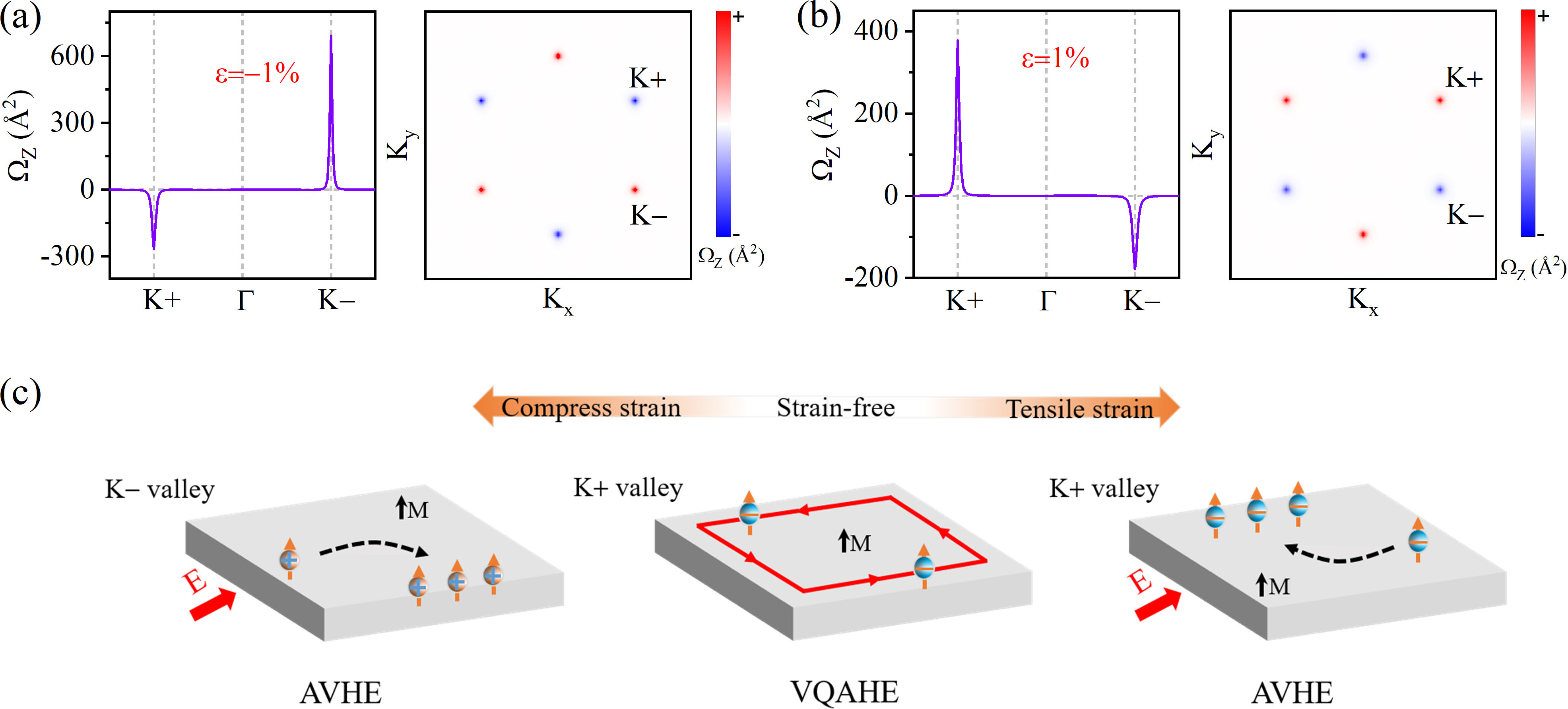}
	\caption{(a) and (b) The distribution of Berry curvature along the high symmetry path and along the whole BZ under $\varepsilon=-1\%$ and $\varepsilon=1\%$, respectively. (c) Schematic diagram of multiple valley Hall effect in ML VSiGeN$_4$ tuned by strain. The holes and electrons are denoted by blue “$+$” and orange “$-$” symbol, respectively. Upward arrows in orange color refer to the spin-up carriers. Red frame represents edge state.}
	\label{fgr:5.jpg}
\end{figure*}

As shown in Fig.~\ref{fgr:4.jpg}(a), from $\varepsilon=-$1$\%$ to $\varepsilon=$ 1$\%$, the orbital composition of the band at two valleys is reversed. This band inversion occurs sequentially in the K$-$ ($\varepsilon=-$0.32$\%$) and K$+$ valleys ($\varepsilon=$ 0.1$\%$) via two HVM states, implying the occurrence of topological phase transition. We calculated edge state of the material under $\varepsilon=\pm$1$\%$, respectively. As shown in Fig.~\ref{fgr:4.jpg}(b) and Fig.~\ref{fgr:4.jpg}(c), one clearly observes that an edge state connecting conduction bands ($\varepsilon=-1\%$) or valence bands ($\varepsilon=1\%$). Therefore, the system transforms from topological nontrivial state (VQAH state) to trivial state (FVS) under strain. In order to understand the division of different phases more intuitively, we plot a phase diagram with the variation of the band gaps at two valleys for ML VSiGeN$_4$. As shown in Fig.~\ref{fgr:4.jpg}(d), the band gaps change linearly with strain at two valleys, and the band gap at each valley has a process of closing and reopening, where the gapless states in the process correspond to the two HVM states. During this process, a topological phase transition from VQAH state to FVS occurs. Therefore, small strain can induce phase transition and effectively tune the electronic structure along with the valley state. Besides, we also calculated energy difference between different magnetic states under strain (Fig.~S3), which reveal that FM state is still the ground state under different strain.

For FVS systems, there exist nonzero Berry curvature with different values at K$+$ and K$-$ valley. As shown in Fig.~\ref{fgr:5.jpg}(a) and Fig.~\ref{fgr:5.jpg}(b), when $\varepsilon=-1\%$, the $\Omega_z(k)$ have different magnitudes with opposite signs at two valleys, which revealing the typical valley contrasting characteristic in ML VSiGeN$_4$. Due to the topological phase transition, both the sign and magnitude of $\Omega_z(k)$ are modified (see Fig.~\ref{fgr:3.jpg}(a)). So, when $\varepsilon=$ 1$\%$, the sign of $\Omega_z(k)$ at K$+$ and K$-$ valley is opposite to the case of $\varepsilon=-$1$\%$. As a result, the sign and magnitude of $\Omega_z(k)$ for different valleys can be changed by applying a small strain. The tunable sign of Berry curvature has been reported by reversing ferroelectric polarization and magnetization \cite{22,35}. Here, we propose a convenient method to modulate the Berry curvature.

With a nonzero $\Omega_z(k)$, the electrons will acquire an anomalous transverse velocity $\upsilon_n(k)$ under an in-plane longitudinal electric field $E$, where $\upsilon_n(k)=\frac{\partial\varepsilon_n(k)}{\hbar\partial k}-\frac{e}{\hbar}E\times\mathrm{\Omega}_n(k)$ \cite{39}, and induce AVHE. Fig.~\ref{fgr:5.jpg}(c) displays multiple valley Hall effects in ML VSiGeN$_4$ which can be tuned by applying strain. For strain-free ML VSiGeN$_4$, the VQAHE with one spin-polarized edge state can be presented. The edge state is chiral-spin-valley locking, which means that when the magnetization direction is reversed, the chiral, spin, and valley of the edge state will also flip. For FVS state under compressive strain (larger than $-$0.32$\%$), when shifting the Fermi level between the K$+$ and K$-$ valleys in the valence band (Fig.~\ref{fgr:4.jpg} and Fig.~S2), the spin-up holes at K$-$ valley flow to right side of the sample under an external in-plane electric field $E$, thus the AVHE occurs. For FVS state under tensile strain (larger than 0.1$\%$), valley polarization occurs at conduction band. In this case, the spin-up electrons from the K$+$ valley accumulated at the left side of the sample under electric field $E$ With spin-up carriers are only accumulated at one side of the sample, the Hall voltage can be detected. Besides, when reverse the magnetization orientation, the conductive valley is changed (from K$+$(K$-$) valley to K$-$(K$+$) valley), and because of the spin flipping, the direction of the lateral movement of the carriers changes accordingly, and thus the sign of Hall voltage changes. Therefore, the valley degree of freedom can be selectively manipulated in ML VSiGeN$_4$ by applying strain and magnetic field.
\begin{figure*}[htbp]
	\centering
	\includegraphics[width=0.85\linewidth]{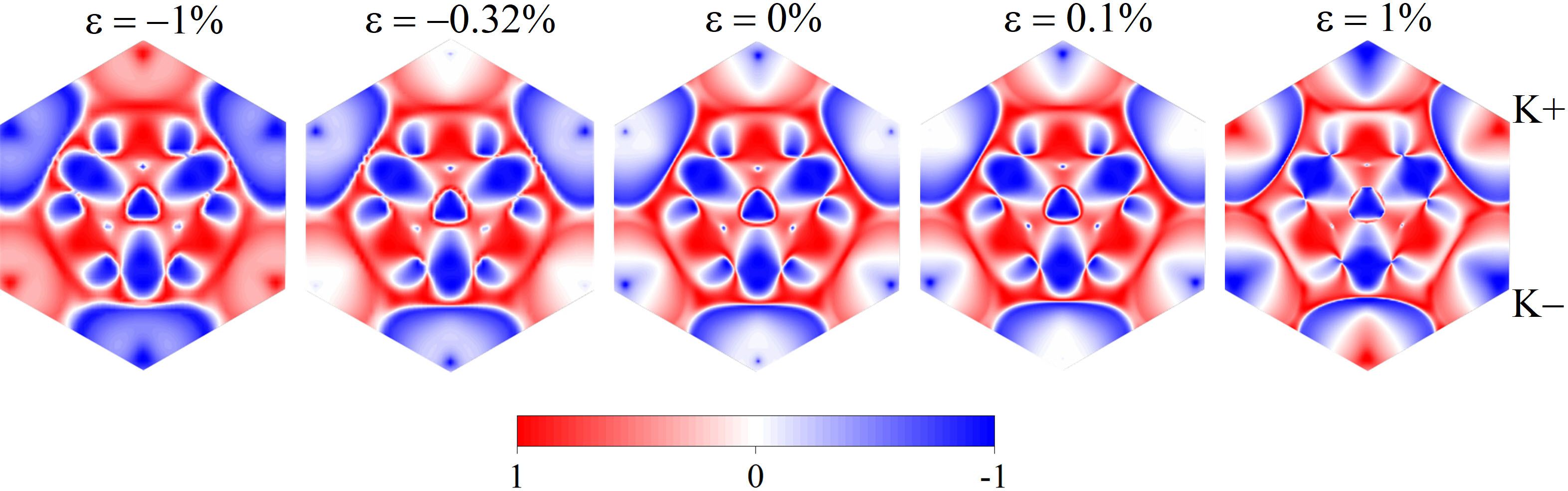}
	\caption{The BZ colour-coded by the degree of circular polarization $\eta(k)$ of MLVSiGeN$_4$ under different strain, respectively. $\pm$1 representative completely absorption left- ($\sigma_+$) and right-handed ($\sigma_-$) polarized light, respectively.}
	\label{fgr:6.jpg}
\end{figure*}

Except for valley-related Hall effect, the different polarized light absorption is also an important topic of valleytronics and valley physics. The chirality of different valleys in the same material is usually different, and one chirality corresponds to one polarized light absorption. The change in sign of $\Omega_z(k)$ under strain implies the change in valley chirality. We calculated valley-selective circular dichroism of ML VSiGeN$_4$ under different strain as displayed in Fig.~\ref{fgr:6.jpg}. The degree of circular polarization $ \eta(k)=\frac{{|M_+(k)|}^2-{|M_-(k)|}^2}{{|M_+(k)|}^2+{|M_-(k)|}^2}$ \cite{8} indicates the difference between the absorption of right-handed and left-handed polarized lights ($\sigma_\mp$) at each k point, and $M_{\pm}(k)$ is the transition matrix of circular polarization. One clearly observes that both the K$\pm$ valleys mainly absorbs $\sigma_-$ in strain-free ML VSiGeN$_4$. This selective absorption of polarized light can be used in optoelectronic devices such as filter. For two HVM states, the material still mainly absorbs $\sigma_-$, but only one of the two valleys has light absorption because the other valley is gapless. Differently, the absorption occurs at K$+$ valley for HVM1 state ($\varepsilon=-0.32\%$) and K$-$ valley for HVM2 state ($\varepsilon=0.1\%$). Under a larger stain, the phase transition will occurs, and the valley chirality will change. As shown in Fig.~\ref{fgr:6.jpg}, when $\varepsilon=\pm1\%$, the two valleys absorb different polarized lights, and for a particular valley, opposite polarized light is absorbed under tensile and compressive strain. As a result, strain can also induce change of valley chirality, enabling the manipulation of polarized light absorption at different valleys.

\section{CONCLUSIONS}
In summary, combining first principles calculations with Wannier-function-based tight-binding model, we demonstrate the structural stability, ferromagnetism, valley-dependent properties, topological properties and strain-induced phase transition of ML VSiGeN$_4$. The results show that ML VSiGeN$_4$ is a ferromagnetic semiconductor with out-of-plane magnetization and a $T_C$ of 113 K. Simultaneously, it intrinsically exhibits a VQAHE with 100$\%$ spin polarization edge state and spontaneous valley polarization. By applying tensile strain of 0.1$\%$ and compressive strain of 0.32$\%$, the HVM state with 100$\%$ valley polarization can be acquired. The FVS state can be achieved by further increasing the strain, and it has an AVHE that can be modulated by flipping the magnetization direction. The strain-tunable polarized light absorption at different valleys can be applied to the next generation photoelectric device. Our work provides a good material platform for the interplay between valleytronics, spintronics, and topology.

\newpage
\begin{acknowledgments}
This work was financially supported by the National Natural Science Foundation of China (No.~11904079, No.~12104130, No.~12047517), the Natural Science Foundation of Henan (No.~202300410069), the China Postdoctoral Science Foundation (No.~2020M682274, No.~2020TQ0089) and the Foundation of Henan Educational Committee (No.~22A140015). Thanks to Dr.~Shan Guan from Institute of Semiconductors in Chinese Academy of Sciences and Dr.~Botao Fu from College of Physics and Electronic Engineering in Sichuan Normal University for their helpful discussions. The authors acknowledge Beijng PARATERA Tech CO.,Ltd.~for providing HPC resources that have contributed to the research results reported within this paper.
\end{acknowledgments}
\bibliography{Manuscript}

\end{document}